\documentclass[Conference]{IEEEtran}
\usepackage[left=0.63in,right=0.63in,top=0.8in,bottom=1.05in]{geometry}
\usepackage{multirow}
\usepackage{bbold}
\usepackage{ifpdf}
\usepackage{ulem} 
\usepackage{subcaption}%
\usepackage[cmex10]{amsmath}
\usepackage{amssymb}
\usepackage{verbatim}
\usepackage{algorithm}
\usepackage{algorithmic}
\usepackage{array}
\usepackage{latexsym}
\usepackage{color}
\usepackage{textgreek}
\usepackage{subscript}
\usepackage{rotating}
\usepackage{booktabs,multirow}
\usepackage{mdframed}	
\usepackage{tabularx}
\usepackage{amsmath}
\usepackage{makecell}
\usepackage{tabto}
\usepackage{mathrsfs}
\usepackage{url}
\usepackage{cite}
\usepackage{listings}
\usepackage{array}
\usepackage[table]{xcolor}
\usepackage{ragged2e}
\usepackage{booktabs}
\usepackage{graphicx}

\usepackage[cmex10]{amsmath}
\usepackage{amssymb}
\usepackage{bbold}
\usepackage{float}
\usepackage[cmex10]{amsmath}
\usepackage{amssymb}
\usepackage{verbatim}
\usepackage{array}
\usepackage{latexsym}
\usepackage{color}
\usepackage{tabularx} 
\usepackage{textgreek}
\usepackage{subscript}
\usepackage{rotating}
\usepackage{booktabs,multirow}
\usepackage{mdframed}	
\usepackage{tabularx}
\usepackage{amsmath}
\usepackage{makecell}
\usepackage{tabto}

\usepackage{mathrsfs}
\usepackage{url}
\usepackage{cite}
\usepackage{listings}
\usepackage{array}
\usepackage[table]{xcolor}

\definecolor{pinkpurple}{rgb}{0.6, 0.1, 0.9} 
\usepackage{algorithmic}
\usepackage{array}
\usepackage{soul}
\usepackage{comment}
\sethlcolor{green}
\usepackage{url}
\usepackage{pifont}
\usepackage{xcolor}
\usepackage{multirow}
\usepackage{graphicx} 

\begin{document}

\title{Systems-Level Planning and Coordination of Truck–Drone Collaborative Delivery Networks}
\author{
	Didem Cicek, ~Burak Kantarci,~\IEEEmembership{Senior Member,~IEEE}
	\thanks{
	D. Cicek and B. Kantarci are with the School of Electrical Engineering and 
        Computer Science at the University of Ottawa, Ottawa, ON, K1N 6N5, Canada.
		E-mails: \{dcice028, burak.kantarci\}@uottawa.ca.}
\vspace{-0.3in}
}

\maketitle
\thispagestyle{empty}
\pagestyle{empty}

\begin{abstract}
\setlength{\columnsep}{0.2 in}
Urban last-mile parcel delivery increasingly relies on heterogeneous fleets whose performance depends on timely coordination, reliable communication, and scalable control. Truck–drone collaboration has emerged as a networked cyber-physical delivery paradigm that combines the payload capacity and range efficiency of trucks with the agility of drones in congested or access-limited urban environments. This paper proposes a layered planning and coordination framework that structures truck–drone collaborative delivery (TDCD) from a systems and control perspective. The framework consists of five interrelated layers: spatial–demand alignment, collaborative delivery configuration, resource and workflow orchestration, performance evaluation, and scalability analysis, providing a unified view of coordination, control, and system-level performance in networked delivery operations. The proposed framework is evaluated using a realistic urban last-mile delivery scenario derived from the 2021 Amazon Last Mile Routing Research Challenge dataset. The case study demonstrates how coordinated truck–drone operation, enabled by structured task orchestration and inter-agent synchronization, improves end-to-end system efficiency under operational constraints. Results show a 42.4\% reduction in total delivery time and a 44.2\% reduction in energy consumption compared to a conventional truck-only delivery model. The scalability analysis further highlights how coordination gains persist as system size increases, and shows the importance of efficient control and communication in heterogeneous delivery networks.
\end{abstract}

\begin{IEEEkeywords}
\setlength{\columnsep}{0.2 in}
Last-Mile Innovation, Sustainable delivery, Truck-Drone Collaboration, Conceptual Framework, TDCD \end{IEEEkeywords}
\IEEEpeerreviewmaketitle

\section{Introduction}
The rapid growth of e-commerce and mounting concerns over traffic congestion, emissions, and heightened service expectations have led to the introduction of unmanned aerial vehicles (UAVs), commonly referred to as delivery drones, as a potential complement to conventional truck-based logistics. Nevertheless, the widespread integration of drones into urban delivery networks remains limited by operational and regulatory constraints. Consequently, recent scholarship has increasingly turned toward hybrid truck–drone delivery systems, wherein drones function in coordination with ground vehicles rather than as independent agents.

Existing research has diverse frameworks to examine truck–drone collaboration. Collectively, these studies demonstrate that collaborative truck–drone configurations can outperform truck-only delivery, yielding reductions in delivery time, operational costs, and environmental impacts. However, no unified planning framework has yet emerged to systematically integrate these approaches and provide a road-map to scale these operations beyond pilot implementations.

\begin{figure*}[!hbt]
        \centering
        \includegraphics[width = 0.7\textwidth, trim=0cm 0cm 0cm 0cm,clip]{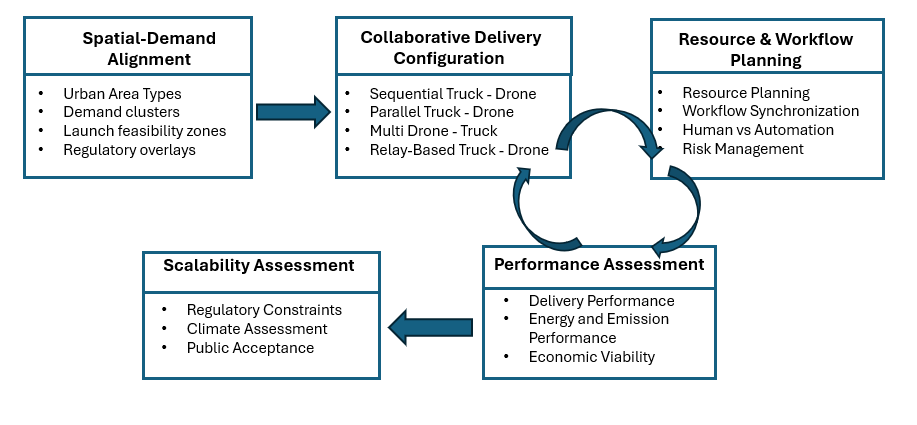}
        \caption{ Planning Framework for Scaling Truck-Drone Collaborative Delivery }
        \label{fig:framework}
\end{figure*}

This paper introduces a multi-layer planning framework for truck–drone collaborative delivery (\figurename\ref{fig:framework}). The framework explicitly differentiates between spatial readiness, collaboration design, operational coordination, performance evaluation, and scalability assessment. By synthesizing insights across regulatory, methodological, and operational domains, this study demonstrates the proposed framework through a case study based on a real-world routing dataset, evaluating the delivery performance of a truck–drone collaborative model for scalable urban last-mile parcel delivery.

This paper is structured as follows: 
Section \ref{sec:litreview} discusses the related work and highlights the novelty of this framework. Section \ref{sec:framework} lays out the planning framework explaining the interdependent collaborative layers of the framework. A case study to demonstrate the framework is covered in Section \ref{sec:casestudy}. Finally, Section \ref{sec:conc} presents the concluding remarks for this work with directions for future research.

\section{Related Work}
\label{sec:litreview}

Existing research on truck–drone collaborative delivery has predominantly focused on optimization-centric formulations, where the primary objective is to minimize delivery time, distance, or energy consumption under predefined operational assumptions. The majority of proposed frameworks adopt problem-specific mathematical models such as mixed-integer linear programming, heuristics, or metaheuristics tailored to a single collaboration mode, static demand conditions, or fixed fleet configurations. While these studies offer valuable algorithmic insights, they typically treat planning decisions, operational execution, and performance evaluation in isolation, with limited consideration of end-to-end system design, workflow coordination, or scalability under real-world constraints. In contrast, the planning framework proposed in this paper adopts a holistic, multi-layered perspective that integrates collaboration typologies, resource and workflow planning, performance assessment, and scalability considerations within a unified structure. As summarized in the comparative table \tablename \ref{tab:Relatedwork}, the proposed framework extends beyond optimization, addresses key limitations of existing approaches and enabling more transferable, decision-oriented insights for real-world deployment. 
    
\begin{table*}[]
\caption{Comparative Table: Existing Frameworks vs. Proposed Framework}
\label{tab:Relatedwork}
\begin{tabular}{|l|l|l|l|l|}
\hline
\textbf{Framework / Study}                                                 & \textbf{Primary Focus}                                                                          & \textbf{Framework Type}      & \textbf{Scalability Treatment}                                                       & \textbf{Key Limitations}                                                                        \\ \hline
\textbf{\cite{DOOLE2020101862}}                                               & \begin{tabular}[c]{@{}l@{}} Drone-based package delivery \\ traffic density estimation \end{tabular}             & Methodological               & \begin{tabular}[c]{@{}l@{}}Indirect (airspace capacity \\ planning)\end{tabular}     & \begin{tabular}[c]{@{}l@{}}No truck–drone interaction\end{tabular}    \\ \hline

\textbf{\cite{she2023analysis}}                                         & \begin{tabular}[c]{@{}l@{}} UAV supported last-mile \\ delivery systems\end{tabular}          & Operational and optimization & \begin{tabular}[c]{@{}l@{}}Explicit computational \\ scalability limits\end{tabular} & \begin{tabular}[c]{@{}l@{}}Optimization becomes \\ intractable at scale\end{tabular}            \\ \hline
\textbf{\cite{electronics14102026}}                                               & \begin{tabular}[c]{@{}l@{}}ML-driven truck–drone\\  resource decisions\end{tabular}             & Operational and optimization & \begin{tabular}[c]{@{}l@{}}Scenario-based scalability \\ analysis\end{tabular}       & \begin{tabular}[c]{@{}l@{}}Framework focused on \\ operational layer\end{tabular}               \\ \hline
\textbf{\cite{bao2025future}}                                                 & \begin{tabular}[c]{@{}l@{}}Environmental and economic \\ sustainability\end{tabular}            & Operational and optimization & Sensitivity-based scalability                                                        & \begin{tabular}[c]{@{}l@{}}Limited collaboration\\  taxonomy\end{tabular}                       \\ \hline
\textbf{\cite{mahmoodi2025framework}}                                                     & Hybrid RPAS–truck optimization                                                                  & Operational and optimization & Explicit scalability evaluation                                                      & Algorithm-centric                                                                               \\ \hline
\textbf{\cite{li2023aviation}}                                               & \begin{tabular}[c]{@{}l@{}}Traffic Management and \\  resource allocation \end{tabular}                                                                                               & Batch Optimization        & Computational scalability                                                               & More aviation oriented                                                                          \\ \hline
\textbf{\cite{lakhwani2025integrating}}                                         & \begin{tabular}[c]{@{}l@{}}Embedding drone deliveries \\ in 5PL systems\end{tabular}                 & Hybrid policy–operational    & System-level scalability vision                                                      & \begin{tabular}[c]{@{}l@{}}High implementation \\ complexity\end{tabular}                       \\ \hline
\begin{tabular}[c]{@{}l@{}}Proposed Framework \\ (This Paper)\end{tabular} & \begin{tabular}[c]{@{}l@{}}Truck–drone collaboration \\ design and scaling\end{tabular}         & Planning                     & \begin{tabular}[c]{@{}l@{}}Explicit scalability \\ assessment layer\end{tabular}     & \begin{tabular}[c]{@{}l@{}}Conceptual (intended to \\ guide future implementation)\end{tabular} \\ \hline
\end{tabular}
\end{table*}

\section{Planning Framework} 
\label{sec:framework}
Urban last-mile drone delivery can scale only if drones operate collaboratively with trucks, which serve as mobile micro-depots, energy/charging hubs, and logistics synchronizers.
The proposed framework organizes this problem into five interdependent collaborative layers: \ref{sec:spatial} Spatial–Demand Alignment, \ref{sec:collaborative} Collaborative Delivery Configuration, \ref{sec:coordination} Resource and Workflow Planning , \ref{sec:performance} Performance Assessment  and \ref{sec:scalability} Scalability Assessment . Each layer builds on the previous one. Rather than assuming a strictly sequential process, the framework incorporates a circular feedback loop between the collaborative delivery configuration, resource and workflow planning, and performance assessment layers, recognizing that operational decisions and resource allocations must be iteratively refined in response to performance outcomes and evolving system conditions. The following sections describe each framework layer, outlining its scope and key analytical dimensions.

\subsection{Spatial-Demand Alignment} 
\label{sec:spatial}
This layer aims to understand the operational complexity involved and to identify where collaborative drone operations are most valuable. Urban environments differ in density, accessibility, traffic congestion, and flight restrictions. Scaling drones requires matching spatial opportunity with operational feasibility. Key analytical dimensions include: 
\textit{Urban area types}: The delivery area is examined to identify spatial characteristics and is classified according to urban morphology, such as high-rise cores, mixed-use corridors, residential suburbs, and cul-de-sacs.
\textit{Demand clusters}: Geographic groupings of customer demand are identified through analysis of order data to assess where drone deployment can most effectively reduce truck delivery workloads, as performed in our previous work \cite{cicek2024green}.
\textit{Launch feasibility zones}: Areas are examined to determine where trucks can safely stop and drones can take off and land.
\textit{Regulatory overlays}: Regulatory and policy constraints are overlaid onto the service area to delineate no-fly zones, restricted airspace around sensitive facilities, and privacy-related limitations governing drone operations.

\subsection{Collaborative Delivery Configuration} 
\label{sec:collaborative}
While the literature presents numerous operational variants of truck–drone delivery, a consistent taxonomy remains absent. We propose four fundamental truck–drone collaborative models that capture the essential modes of synchronization, fleet scaling, and asset coupling observed in practice. More complex or adaptive strategies can be viewed as hybrid combinations of these baseline collaboration types.

\subsubsection{\textbf{Sequential Truck-Drone Collaboration}} 
The truck and drone operate in a tightly synchronized manner, where the truck temporarily stops and waits for the drone to complete its delivery and return. During this time, the driver may perform auxiliary tasks such as parcel sorting or preparation. This corresponds closely to  Traveling Salesman Problem with Drone (TSP-D) \cite{DING2024100025}, \cite{gunay2023hybrid} where the truck remains stationary at a parking node while the drone visits assigned customers and returns to the truck. 
This type of collaboration can help delivery in narrow streets or restricted access areas.

\subsubsection{\textbf{Parallel Truck-Drone Collaboration}} 
The truck continues along its delivery route while the drone performs a delivery independently and rejoins the truck at a later location. Truck and drone operations are partially decoupled in time. This configuration also aligns with TSP-D where truck remains active instead of waiting for the drone to complete its delivery. Parallel Drone Scheduling TSP (PDSTSP) \cite{montemanni2023solving}, and Flying Sidekick TSP (FSTSP) \cite{DING2024100025}, \cite{MURRAY201586} in the literature try to address the same problem. This type of collaboration mostly fits for suburban neighborhoods with moderate parcel density and predictable road networks. 

\subsubsection{\textbf{Multi-Drone Truck Collaboration}} 
A single truck functions as a mobile logistics hub capable of launching and recovering multiple drones, enabling simultaneous drone deliveries within a localized service area. This mode increases throughput but requires coordination among multiple airborne assets. This collaboration type has been referred in the literature as Multiple Flying Sidekick TSP (mFSTSP) \cite{dell2021modeling}, Flexible Drones TSP (FDTSP) \cite{LU2022118351} or TSP with multiple drones (TSP-mD) \cite{tu2018traveling} model. This collaboration type can serve high parcel density clusters or peak-demand periods.

\subsubsection{\textbf{Relay-Based Truck-Drone Collaboration}}
Delivery tasks are completed through staged hand-offs among multiple trucks and drones. A drone may be launched from one truck and recovered by another, enabling extended operational reach beyond individual drone range limitations. This delivery type has the most operational complexity, hence studied the least by the scholars. Previous research on this problem can be listed as \cite{GAO2023104407}, \cite{WANG2019350}. This delivery model can be used in large or spatially fragmented urban areas, or networks with strict drone range constraints.

\subsection{Resource and Workflow Planning} 
\label{sec:coordination}
This layer defines how truck and drone resources coordinate to scale operations. Key analytical dimensions include: 
\textit{Resource Planning:} Fleet composition and operational resource parameters are defined, including the number of drones and trucks, drone-to-truck allocation ratios, drone rotation strategies (e.g., charging cycles or battery-swap modules), and truck dwell-time policies. \textit{Workflow Synchronization:} Operational workflows governing truck–drone interactions are specified, including launch and recovery protocols, package handover procedures, queuing arrangements between trucks and drones, supporting communication and positioning systems. \textit{Human vs Automation role allocation:} The division of responsibilities between human operators and automated systems is specified. 
\textit{Risk Management:} Risk mitigation measures ensure reliable and safe operations under uncertain conditions, including procedures for adverse weather (e.g., wind, precipitation, snow), compliance with urban safety constraints, and robust navigation capabilities.

\subsection{Performance Assessment} 
\label{sec:performance}
The objective of this layer is to evaluate the outcomes of alternative truck–drone collaborative delivery configurations across key performance metrics, enabling systematic comparison of operational effectiveness, environmental impact, and economic feasibility. Key analytical dimensions with the proposed metrics include:
\textit{Delivery Performance:} Parcel throughput (the number of parcels delivered collaboratively over a given period or route), delivery time reduction relative to baseline, truck-only operations, and on-time delivery rate are important metrics to watch delivery performance. \textit{Energy and Emission Performance:} Drone energy use per parcel, comparative emissions versus truck-only route and balancing the energy use during peak periods are the key metrics for this dimension. 
\textit{Economic Viability:} From a capital cost and life-cycle perspective, delivery drones present a substantially lower upfront investment than electric delivery trucks often by an order of magnitude, yet their economic lifespan is predominantly constrained by battery degradation, which typically occurs on a much shorter replacement cycle than that of electric truck batteries, thereby necessitating a life-cycle-based rather than purchase-cost-based comparison of fleet assets. 

\subsection{Scalability Assessment} 
\label{sec:scalability}
This layer evaluates whether the collaborative model can scale and under what conditions. Key analytical dimensions include:
\textit{Regulatory Constraints:} Regulatory constraints capture the extent to which existing and evolving legal, aviation, and municipal frameworks enable or limit large-scale deployment of truck–drone collaborative delivery systems. 
\textit{Climate Assessment:} Climate assessment evaluates the suitability and resilience of collaborative delivery operations under region-specific and seasonally varying climatic conditions. Unlike controlled pilot environments, large-scale deployment must contend with persistent exposure to weather variability and climate-related extremes. \textit{Public Acceptance:} Public acceptance reflects the perceived acceptability of widespread truck–drone delivery operations among stakeholders. Even when systems are technically viable and legally permitted, insufficient public support can significantly hinder large-scale adoption.

\section{Case Study}
\label{sec:casestudy}
\begin{figure}[!hbt]
        \centering
        \includegraphics[width = 0.25\textwidth, trim=0cm 0cm 0cm 0cm,clip]{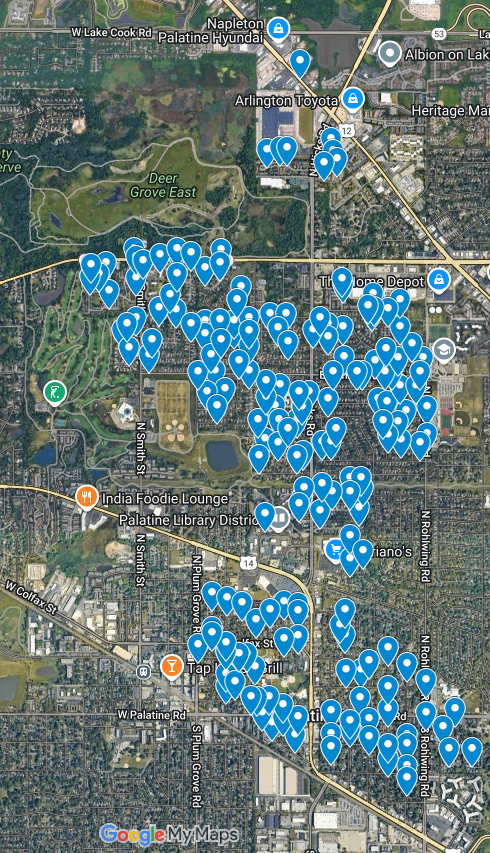}
        \caption{Delivery Area used for the case study}
        \label{fig:route1}
\end{figure}

The proposed framework is applied to a truck–drone collaborative delivery scenario using a single delivery route selected from the 2021 Amazon Last Mile Routing Research Challenge \cite{merchan20242021}. The dataset released for the Amazon Last Mile Routing Research Challenge contains route-, stop-, and package-level attributes for 9,184 historical delivery routes executed by Amazon drivers in 2018 across five major metropolitan areas in the United States. The case study in this paper focuses on a single route in Palatine, Illinois, located within the Chicago metropolitan area, comprising 192 customer locations and spanning an operational service area of approximately 18 square kilometers (RouteID\textsubscript{15baae2d-bf07-4967-956a-173d4036613f}). The delivery area as mapped on GoogleMyMaps\footnotemark{}   \footnotetext{https://www.google.com/maps/about/mymaps/} can be found on \figurename\ref{fig:route1}. The delivery truck is assumed to be a Battery Electric Vehicle (BEV) with a driver onboard. The objective of this demonstration is to operationalize each framework layer by outlining the key steps that support informed decision-making in the design of scalable, efficient, and economically viable TDCD models for urban last-mile delivery.

\subsection{Layer 1: Spatial-Demand Alignment}
\begin{figure}[!hbt]
        \centering
        \includegraphics[width = 0.40\textwidth, trim=0cm 0cm 0cm 0cm,clip]{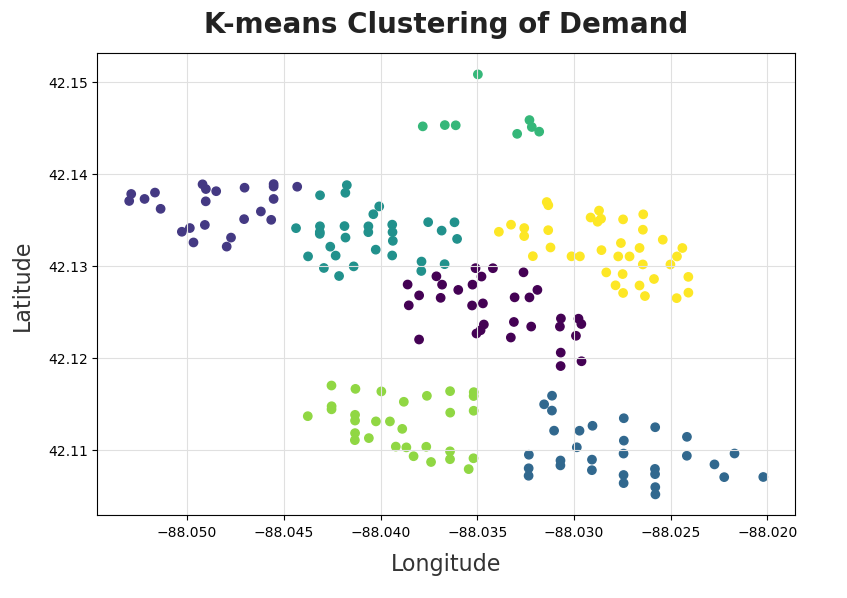}
        \caption{Clusters of Delivery Demand}
        \label{fig:clusters}
\end{figure}

\begin{figure}[!hbt]
        \centering
        \includegraphics[width = 0.40\textwidth, trim=0cm 0cm 0cm 0cm,clip]{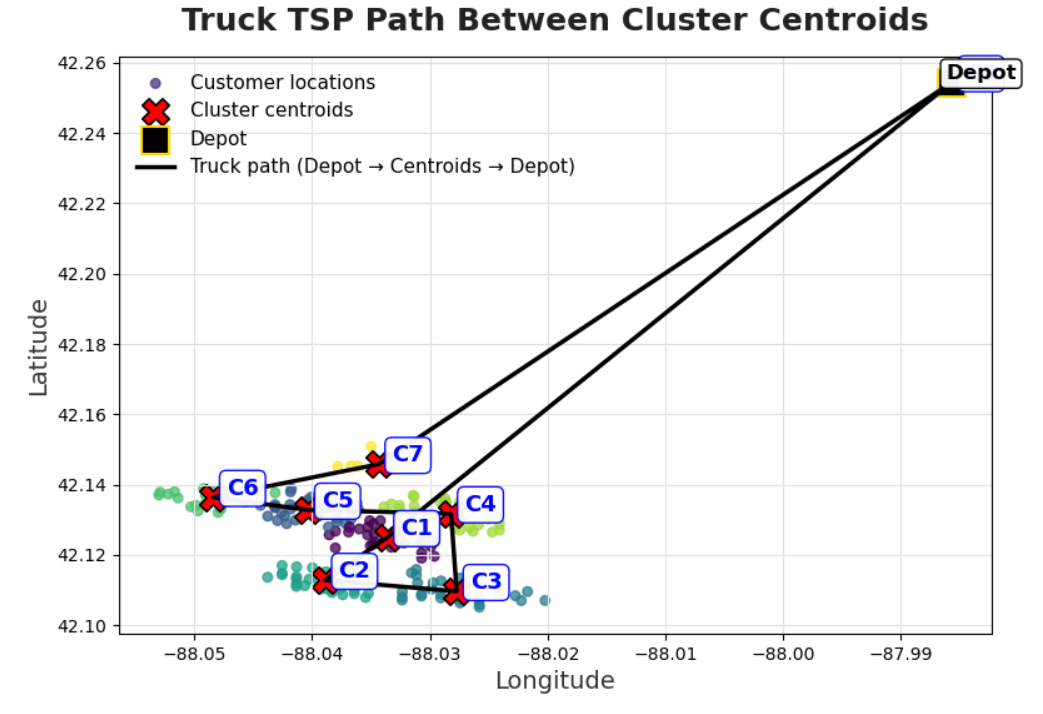}
        \caption{Truck TSP Path and Potential Truck Stops}
        \label{fig:truckstops}
\end{figure}

The selected delivery area is a residential suburban fabric dominated by single-family neighborhoods, local streets, schools and churches as seen on the aerial imagery. Due to relatively predictable curb environments and no-hard parking restrictions, the area is good to scale truck-drone collaborative delivery. Although drone operations must comply with Federal Aviation Administration (FAA) regulations, local and district-level rules must also be satisfied prior to deployment. Customer demand, assumed to be known in advance, is then analyzed to identify spatial groupings using proximity-based features. The K-means clustering algorithm is applied to the dataset, resulting in seven distinct demand clusters, as shown in \figurename\ref{fig:clusters}. Each cluster consists of 30 customers on average, except for one cluster (C7) with 8 customers. Then the centroids of each cluster is calculated as a guidance for selecting potential truck stops to launch the drone \figurename\ref{fig:truckstops}. The centroid is just a mathematical point and may not directly translate into an actual location; further investigation of the location is required considering the building and road infrastructures.

\subsection{Layer 2: Collaborative Delivery Configuration}
The selection of a truck drone collaborative delivery configuration is governed by the interaction between key decision variables including delivery fleet scale reflecting operator resource capacity, feasible launch and recovery locations along the truck route, and customer allocation between truck and drone service, and a set of operational constraints encompassing drone energy limits, flight range, truck and drone synchronization and safety requirements, and regulatory restrictions such as controlled airspace and permit required zones. Given the clustered suburban morphology of the Palatine case area and the close spatial proximity of customers within each identified demand cluster, a multi-drone truck collaboration operating mode is selected as the baseline configuration to minimize delivery completion time.

\subsection{Layer 3: Resource and Workflow Planning}
At this operational layer, this case study adopts a set of simplifying assumptions to enable a transparent demonstration of coordinated truck–drone operations. The delivery fleet consists of one truck carrying three drones, representing a moderate but realistic operational scale for a suburban service area. Each drone is assumed to carry one parcel per sortie, with no mid-flight handoffs or batch deliveries. Drone batteries are assumed to be swapped on the truck by the driver, implying human-in-the-loop ground operations and eliminating the need for automated charging or robotic handling infrastructure. Truck is acting mainly as a mobile depot and not used for parallel delivery. Weather conditions are assumed to be favorable, with no wind, precipitation, or visibility constraints affecting flight performance. All drones are modeled as DJI Matrice 100 equipped with TB47D battery \footnotemark{}   \footnotetext{https://www.dji.com/ca/support/product/matrice100},with a payload of 500 gr and a constant horizontal flight speed of 8m/sec over the Euclidean distance at an altitude of 40 meters. The truck is modeled as a Fiat Ducatio BEV 3.5t \cite{balassa2023sustainability} and assumed to operate at an average cruising speed of 30 km/h on suburban residential roads.

\subsection{Layer 4: Performance Assessment}
Delivery performance of a baseline truck-only route, formulated as a Traveling Salesman Problem (TSP) and solved using a greedy heuristic is compared against a multi-drone truck collaborative delivery scenario  in terms of delivery time and energy considerations. Amazon Routing Dataset contains daily routes of average 150-200 stops. The total delivery time for the truck-only scenario on the selected route, comprising 192 customer locations, is reported as 204.7 minutes for transit, 222.7 minutes for package service (i.e handling, delivery to doorstep), 24.4 minutes for return to depot, amounting to total 451.8 minutes (7.5 hr) delivery time. In the multi-drone truck–drone collaboration scenario, total delivery time is computed as the sum of the truck’s travel time between cluster centers (using exact TSP), drone loading time, drone flight time based on Euclidean distances between cluster centroids and customer locations, and vertical ascent and descent times. The simulation has been run on Python (v 3.10.18) and the results show that collaboration model led to 42.4\% reduction in total delivery time \tablename \ref{tab:Delivery_time}. It is important to note that the magnitude of the reported savings is highly sensitive to the assumptions adopted in this study regarding cruising altitude, drone speed, payload, and parcel loading time.

\begin{table}[]
\caption{Delivery time comparison of Multi-Drone Truck Collaboration vs Truck-only Delivery}
\label{tab:Delivery_time}
\begin{tabular}{|l|c|c|}
\hline
\textbf{}                                                  & \textbf{Truck only} & \textbf{Multi-Drone Truck}               \\ \hline
\textbf{Truck travel   distance-km}                        & 63.10            & 20.93                                  \\ \hline
\textbf{Truck   speed-km/hr}                               & 30             & 30                                  \\ \hline
\textbf{Drone travel   distance-km}                        & -                   &161.5                                 \\ \hline
\textbf{Drone cruise speed (m/s)}                              & -                   & 8                                  \\ \hline
\textbf{Drone   ascending speed (m/s)}                           & -                   & 3                                   \\ \hline
\textbf{Drone   descending speed (m/s)}                          & -                   & 2.5                                   \\ \hline
\textbf{Drone cruising   altitude (m)}                         & -                   & 40                                    \\ \hline
\textbf{Drone handling   time (sec)}                              & -                   & 30                                   \\ \hline
\textbf{Total delivery   time (min)}                             & \textbf{451.80}    & \textbf{{260.3}}                       \\ \hline
\cellcolor[HTML]{FBE2D5}\textbf{Delivery time   reduction} & -                   & \cellcolor[HTML]{FBE2D5}\textbf{-42.4\%} \\ \hline
\cellcolor[HTML]{FBE2D5}\textbf{Delivery time/parcel (min)} & 2.4                   & \cellcolor[HTML]{FBE2D5}\textbf{1.4} \\ \hline
\end{tabular}
\end{table}

For drone energy calculations, we use a hover endurance of 19.5 min (average of no-payload and 0.5 kg payload hover specifications to reflect loaded outbound and unloaded return legs) and assume an 30\% battery degradation factor to conservatively account for mission power variation across forward flight and vertical phases. This conservatism is motivated by prior endurance models showing speed-dependent power in forward flight and different power requirements in ascent/descent relative to hover \cite{hwang2018practical}. The simulation shows 44.2\% saving in total vehicle energy given the parameters as define on \tablename \ref{tab:Energy}. Vertical energy consumption is modeled as time-based power usage with operational vertical speeds of 3 m/s for ascent and 2.5 m/s for descent, selected conservatively below manufacturer limits to reflect realistic delivery conditions. The lower descent speed results in a longer descent duration and therefore higher energy consumption per unit vertical distance when expressed in Wh/km. 

As both vehicles are electrically powered, tailpipe emissions are absent; however, upstream emissions arise from the electricity generation mix. A comparative assessment of CO2 emissions between a Fiat BEV truck and a DJI Matrice 300 RTK drone is reported in \cite{balassa2023sustainability}, indicating that the drone produces approximately 17\% of the truck’s emissions.

\begin{table}[]
\caption{Energy consumption comparison of Multi-Drone Truck Collaboration vs Truck-only Delivery}
\label{tab:Energy}
\begin{tabular}{|l|l|c|}
\hline
\textbf{}                                               & \multicolumn{1}{c|}{\textbf{Truck only}} & \textbf{\begin{tabular}[c]{@{}c@{}}Multi-Drone \&\\  Truck\end{tabular}} \\ \hline
\textbf{Drone   Battery-Model}                          &                                          & TB47D                                                                    \\ \hline
\textbf{Drone   Battery-Energy}                         &                                          & 99.9 Wh                                                                  \\ \hline
\textbf{Hovering   Time-No payload}                     &                                          & 22 min                                                                   \\ \hline
\textbf{Hovering   Time-500gr payload}                  &                                          & 17 min                                                                   \\ \hline
\textbf{Hovering   Time-Average}                        &                                          & 19.5 min                                                                 \\ \hline
\textbf{Battery Degradation Factor}                        &                                          & 30\%                                                                 \\ \hline\textbf{Max cruise range at 8m/s}                                   &                                          & 6.55 km                                                                  \\ \hline
\textbf{Drone Energy at cruise}                     &                                          & 15.25 Wh/km                                                              \\ \hline
\textbf{Drone Energy at ascend}                     &                                          & 40.66 Wh/km                                                              \\ \hline
\textbf{Drone Energy at descend}                     &                                          & 48.79 Wh/km                                                              \\ \hline
\textbf{Drone Total Energy}                           & \textbf{}                                & \textbf{3.37 k WH}                                                       \\ \hline
\textbf{Truck Energy   consumption}                     & \multicolumn{1}{c|}{235.25 Wh/km}        & 235.25 Wh/km                                                             \\ \hline
\textbf{Truck Total   Energy}                           & \multicolumn{1}{c|}{14.845 k Wh}  & 4.92 k Wh                                                      \\ \hline
\cellcolor[HTML]{FBE2D5}\textbf{Total Vehicle   Energy} & \multicolumn{1}{c|}{14.845 k Wh}           & \cellcolor[HTML]{FBE2D5}\textbf{8.29 k Wh }                          \\ \hline
\cellcolor[HTML]{FBE2D5}\textbf{Energy Saving vs Truck 0nly}          & \multicolumn{1}{c|}{-}                   & \cellcolor[HTML]{FBE2D5}\textbf{{-44.2\%}}                                \\ \hline
\cellcolor[HTML]{FBE2D5}\textbf{Energy per parcel} &    77.32 Wh                   & \cellcolor[HTML]{FBE2D5}\textbf{43.18 Wh} \\ \hline
\end{tabular}
\end{table}

\subsection{Layer 4: Scalability Assessment}
All customer orders on the selected Amazon route are assumed to be drone-eligible, and customers are assumed to prefer drone delivery when available. However, only 64\% of the parcels fall below the 5 lb payload threshold commonly cited for urban drone delivery operations, and public acceptance of drone delivery services remains uncertain. To examine the sensitivity of system performance to adoption levels, the case study simulates a reduced-demand scenario by randomly removing 50\% of customer orders from the routing dataset. In response, the delivery configuration is re-optimized by reducing the number of clusters from seven to three while preserving comparable customer densities per cluster (approximately 30 customers per cluster) and maintaining a fleet of three drones. Under this reduced-demand scenario, the optimized system achieves a 29.8\% reduction in total delivery time relative to the traditional truck-only delivery model. At the same time, total drone energy consumption decreases by 33.5\% compared with the full-demand scenario. The results indicate that although operational benefits remain significant under partial adoption, greater efficiency gains are realized as demand and user adoption of drone delivery increase.

\section{Conclusion and Future Work}
\label{sec:conc}
This paper has addressed a gap in the truck–drone collaborative delivery (TDCD) networks literature by presenting a layered planning and coordination framework that separates system-level performance evaluation from scalability considerations. This distinction is critical in networked cyber-physical systems, where coordination strategies that perform well at limited scale may not remain effective as fleet size, traffic density, or operational constraints increase.

We have evaluated this framework using a realistic urban last-mile delivery scenario based on an Amazon delivery route. 
The results show significant reductions in total delivery time (up to 42.4\%) and energy consumption (up to 44.2\%). It should be emphasized that the present simulation abstracts from several real-world operational dynamics such as stochastic weather conditions, wind disturbances, and drone aerodynamics incorporates these effects only through a uniform 30\% battery degradation factor; therefore, actual field deployment may result in longer service times, an aspect that will be systematically investigated in future research through higher-fidelity operational modeling and real-world validation.


Our ongoing work focuses on extending the framework toward larger-scale and more realistic deployment scenarios, including environmental uncertainty, battery aging effects, and airspace capacity constraints. 

\vspace{0.1 cm}
 \section*{Acknowledgment}
This work was supported in part by Natural Sciences and Engineering Research Council of Canada (NSERC) DISCOVERY and NSERC CREATE TRAVERSAL Programs and in part by the Ontario Research Fund Research Excellence (ORF-RE) program under grant number RE12-024.
\vspace{-0.5 cm}
\normalem
\bibliographystyle{IEEEtran}

\end{document}